%% file: 6491paper.tex
\documentclass{aa}
\usepackage{natbib}
\usepackage{psfrag}
\usepackage{amssymb}
\usepackage[dvips]{graphicx}
\usepackage{afterpage}
\usepackage[latin1]{inputenc} 

\title{Tracking granules on the Sun's surface and reconstructing
horizontal velocity fields: I. the CST algorithm}

\author{M. Rieutord\inst{1}, T. Roudier\inst{2}, S. Roques\inst{1}, 
and C.  Ducottet\inst{3}}

\date{\today}
\offprints{M. Rieutord}

\institute{
Laboratoire d'Astrophysique de Toulouse et Tarbes,
UMR 5572, CNRS et Université Paul Sabatier, 14 avenue E. Belin, 31400
Toulouse,
France
\and Laboratoire d'Astrophysique de Toulouse et Tarbes, UMR
5572, CNRS et Université Paul Sabatier, 57 Avenue d'Azereix, BP
826, 65008 Tarbes Cedex, France
\and Laboratoire Hubert Curien, UMR5516, CNRS et Université Jean
Monnet, 18 rue du Prof. B. Lauras, 42000 Saint-Etienne, France
}

\input 6491com

\begin{document}

\authorrunning{Rieutord et al.}
\titlerunning{The CST algorithm}

\abstract{}{Determination of horizontal velocity fields on the solar
surface is crucial for understanding the dynamics of structures like
mesogranulation or supergranulation or simply
the distribution of magnetic fields.}
{We pursue here the development of a method called CST for coherent
structure tracking, which determines the horizontal motion of granules in
the field of view.}
{We first devise a generalization of Strous method for the segmentation of
images and show that when segmentation follows the shape of
granules more closely, granule tracking is less effective for large granules because
of increased sensitivity to granule fragmentation.  We then introduce
the multi-resolution analysis on the velocity field, based on Daubechies
wavelets, which provides a view of this field on different scales. An
algorithm for computing the field derivatives, like the horizontal
divergence and the vertical vorticity, is also devised. The effects from the
lack of data or from terrestrial atmospheric distortion of the images are
also briefly discussed.}{}

\maketitle

\section{Introduction }

Determining the flows on the surface of the Sun has triggered
many efforts over the last decade. Of special concern was
determination of granules motions that can reveal horizontal flows on
scales larger than, typically, twice their horizontal size $\sim$
2000~km. This scale is small enough to provide observers of the Sun's
surface with a detailed sampling of the large-scale flows, such as
supergranulation, and therefore makes it very interesting to  determine
granule motions.

However, from the viewpoint of fluid mechanics, granules are not passive
scalars whose motions trace that of the fluid; rather, they are
structures in intensity or, assuming perfect correlation with
temperature, coherent structures of temperature. But the evolution of
the temperature field is the result of radiative, as well as advective,
processes. It is only in case that the latter dominates
that the motions of granule can be associated with horizontal flows.
Using numerical simulations in a large horizontal box, we have shown
that  granule motions are highly correlated with horizontal flows {\em when
the scale is larger than $\sim$2500~km} \cite[]{RRLNS01}; below this scale,
granule motions should be considered as (solar) turbulent noise.

Once the equivalence of plasma motion and granule motion is assumed,
one is left with the problem of measuring the latter motion. This is
not an easy task owing to the small angular size ($\sim$ 1.3\arcsec) of
the structures. Ground-based observations are sensitive to atmospheric
turbulence, while space observations are expensive owing to the
(relatively) large aperture needed for resolving granules.

Basically, two techniques have been used to measure horizontal velocity
fields: either the tracking of individual granules \cite[]{Strous95a} or local
correlation tracking \cite[]{NS88}. The results of these two
techniques have been compared \cite[]{SBNSS95,RRMV99} and found to
broadly agree; in the test using numerical simulations \cite[]{RRLNS01},
they show the same degree of correlation with the actual plasma flows.
However, detailed examinations \cite[]{SBNSS95,RRMV99} have demonstrated
worrying differences, especially when field derivatives like vertical
vorticity  and divergence are computed.

In fact from the point of view of signal processing, these two methods
differ fundamentally. On the one hand, granule tracking emphasises
the importance of the granule, gives no signal in between granules, and
yields a velocity field that is sampled randomly following the distribution of
granules. On the other hand, local correlation tracking (LCT hereafter)
treats granules and intergranules on an equal footing and yields a
velocity field on a regular grid. Broadly speaking, the two methods
differ in the interpolation process, which unfortunately influences the
final result.

In this paper we present and analyse in some detail an algorithm
based on granule tracking which is able to give a
reconstruction of the velocity field at all scales larger than the
sampling scale. This algorithm has already been introduced in \cite{RRMV99}
in a preliminary version. We call it CST for coherent structure tracking
to underline its relation with the physics lying behind it. Such an
algorithm is close in its principles to particle-imaging velocimetry
(PIV), as used in experimental fluid mechanics \cite[e.g.][]{Adrian05}.

We developed this algorithm for three reasons: the first is obviously
because it gives a different view of the data than does the LCT algorithm,
since many interpolation problems may influence the final results (see
the discussion in \citealt{PBD03}). The
second one is that it may be used on raw data and gives an estimate of
the error introduced by atmospheric distortion (see the companion paper,
\citealt{TRMR06}, in which this point is developed). Finally, it offers the
possibility of selecting specific structures according to their nature,
size, lifetime, etc. and of studying their motion.

In the next section we discuss the different steps of the algorithm,
especially the segmentation and interpolation processes, and also point
out the effects of regions lacking data. Discussion and conclusions follow.

\section{The CST algorithm}

Before describing the different steps in some detail, let us recall the
five main steps of this algorithm:

\begin{itemize}
\item segmentation of the image and granule identification
\item measurement of velocities at granule locations
\item reconstruction of the velocity field
\item calculation of field derivatives (like the z-component of the
vorticity and divergence)
\item estimation of the noise.
\end{itemize}

\noindent We now go into detail for each one in turn.

\subsection{Segmentation and granule identification}

To identify a granule one needs a criterion with which to decide whether a
given pixel belongs to a given structure or not. This criterion needs to
be local in order to avoid threshold effects due to large-scale
variations in intensity, which either come from terrestrial
atmospheric effects, solar acoustic waves, or even magnetic fields.

A classical criterion is based on detecting local maxima of the
intensity through the curvature $C=I_{i+1}-I_i-(I_i-I_{i-1})$
\cite[]{Strous95a,RRMV99}. This criterion has the advantage  of being
simple, robust, and therefore quite efficient. However, comparing the
detected patterns with the original image shows that this criterion
underestimates the size of the granules.
It is therefore interesting to know whether this criterion
can be improved. {An objective test of this improvement will be
that the lifetime of the granules is increased}.

\begin{figure}[t]
\centerline{\includegraphics[width=8.7cm]{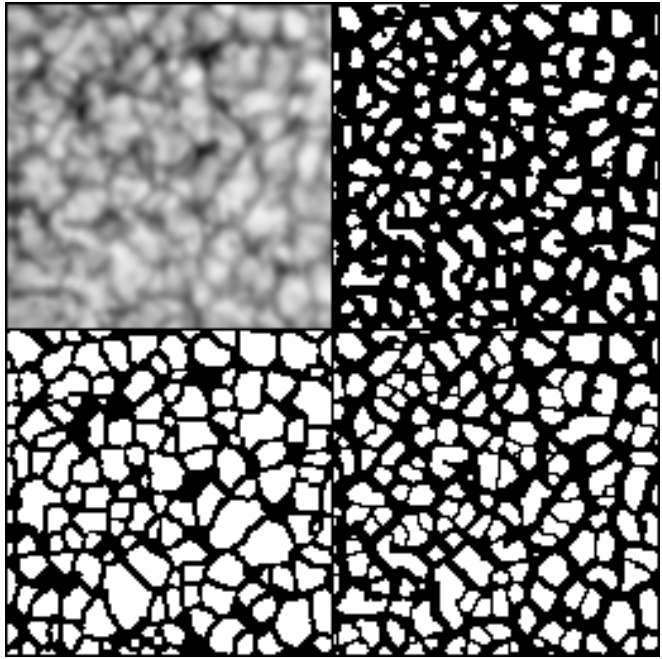}}
\caption[]{Comparison of the result of different segmentations on the
same test image. From top to bottom and left to right we have : the
original image, the result of Strous's method, the result of the BW
method, the result of the proposed method. Note that BW leaves some
granules undetected.} \label{comp_res_segm}
\end{figure}

Another method has been proposed by \cite{BW01} (hereafter referred to as
BW) with a multiple-level tracking algorithm. It is based on the use of
multiple-threshold levels applied to the intensity image.  The granules
detected at a high level are gradually extended to adjacent pixels whose
intensity exceeds the lower level, while keeping a minimum distance with
respect to other granules. This approach, which is very similar to a
watershed-based segmentation \cite[]{VS91,Soille99}, yields sizes and
shapes of granules that conform more to direct observation of the image.
Nevertheless, the number of detected
granules is less compared to the Strous algorithm because this method is based
on the intensity.

The Strous-curvature criterion is more efficient at separating
the granules. It consists in selecting the pixels whose local minimal
curvature is not negative. This curvature is calculated in the four
directions defined by the 8 neighbours of the considered pixel. The
underestimation of the granule size stems from real
granule's extension being greater than the positive curvature region.
We thus propose a new segmentation algorithm (hereafter termed CD)
that combines both ideas
of BW and the Strous algorithm.  It consists in the following steps:

\begin{itemize}
\item Calculation of the ``minimal curvature image'' : for each pixel,
the minimal curvature among the four directions is computed.
\item Detection of the granules as non negative curvature pixels in the
minimal curvature image.
\item Extension of the detected granules with points whose minimal
curvature value is greater than a given (negative) threshold $t_{ext}$,
while keeping a minimal distance of one pixel between each pair of granules.
\end{itemize}

\begin{figure}[t]
\centerline{\includegraphics[width=9.5cm]{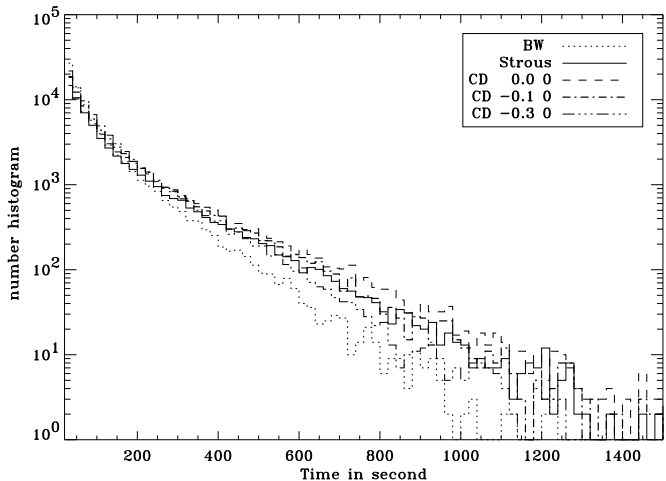}}
\caption[]{Comparison of the effect of different segmentations on
lifetimes of granules. BW refers to the Bovelet and Wiehr method while
CD refers to our method (see text); the associated number is the
extension threshold parameter $t_{ext}$.}
\label{comp_seg}
\end{figure}

\noindent The last step can be reached using the watershed algorithm on the
minimal curvature image, with an additional condition requiring that
the curvature remains above $t_{ext}$.

This new approach leads to a segmentation with the same granules as
Strous's approach, but with a controllable size. Strous's
segmentation is obtained with $t_{ext}=0$. Decreasing the value of
$t_{ext}$ extends the granules. In Fig.~\ref{comp_res_segm} we illustrate
the discussed segmentation method using Pic-du-Midi data\footnote{As for
all examples needing solar images, we used the series obtained at
Pic-du-Midi on 20 September 1988.}. This figure illustrates the way our
segmentation extends that of Strous and closely follows the shape of
granules.

To study the influence of the segmentation on the lifetime of
granules, we plotted the statistic of the lifetime for the three
methods : the Strous method, BW method, and our proposed method.
For our method, we took three different threshold values:
$t_{ext}=0$, $t_{ext}=-0.1$, and $t_{ext}=-0.3$ (Fig.~\ref{comp_seg}).

This figure shows that, broadly speaking, the segmentation does not
influence the statistics of lifetimes very strongly. However, some
differences can be noticed in the detail. The BW algorithm, by detecting
less granules, shows a deficit of short-lived and long-lived granules; on the other
hand, our algorithm eliminates long-lived structures when used with too
low a threshold. We understand this behaviour as the result of enhanced
splitting of larger structures.

In conclusion, Strous's algorithm seems the most efficient for our purpose,
and it can be improved in the way we described, but at the price of
increasing the computation effort a lot.

Once the image has been segmented, each granule needs to be identified.
This operation, although very simple, can be quite time-consuming since
all pixels should be tested at least once. The most efficient way we
have found to deal with  this operation is to use a recursive algorithm,
letting granules grow from a single pixel. A pixel belongs to a
granule if it shares at least one side with another pixel of the granule.

\begin{figure}[t]
\centerline{\includegraphics[width=9cm]{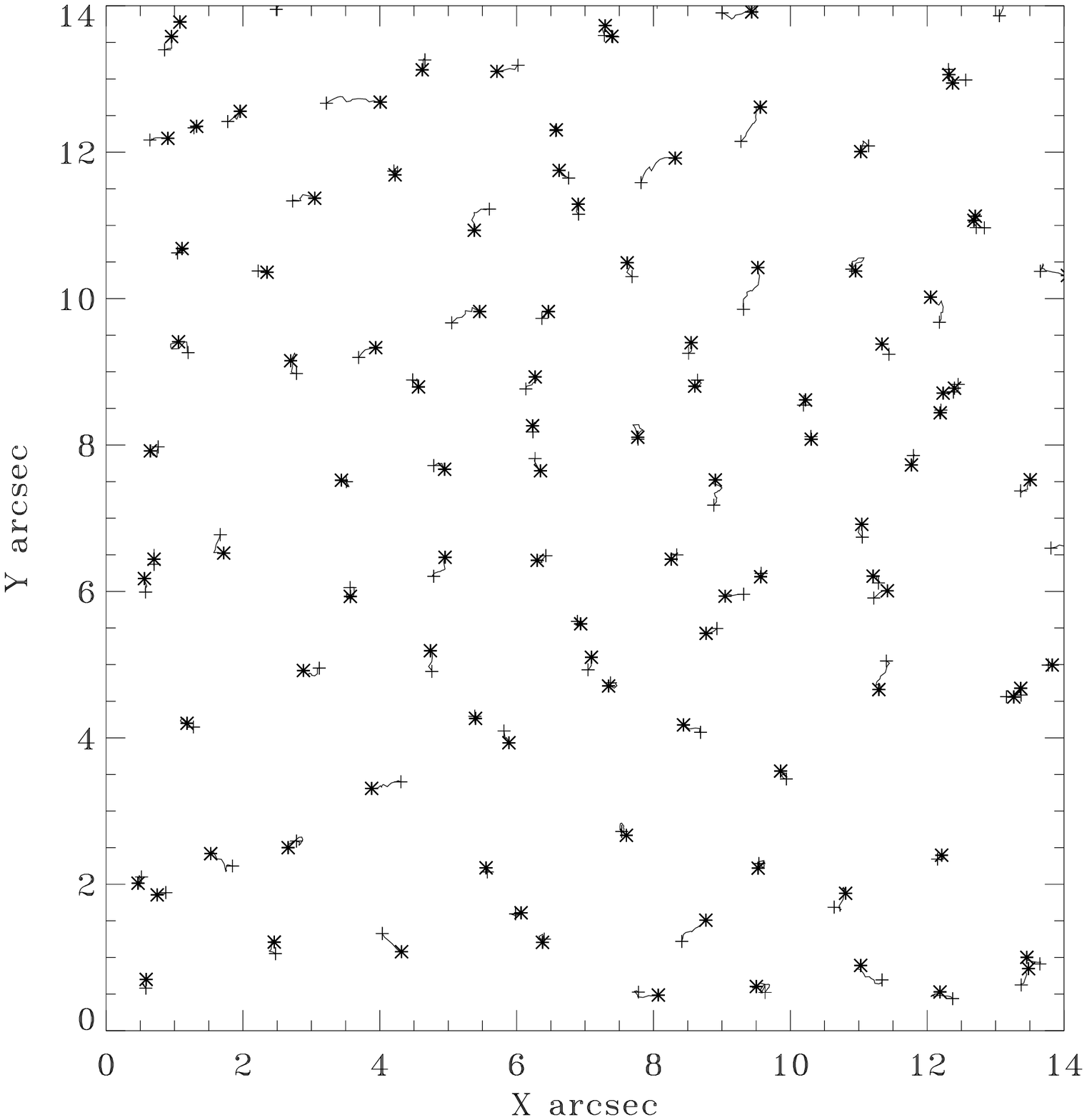}}
\centerline{\includegraphics[width=9cm]{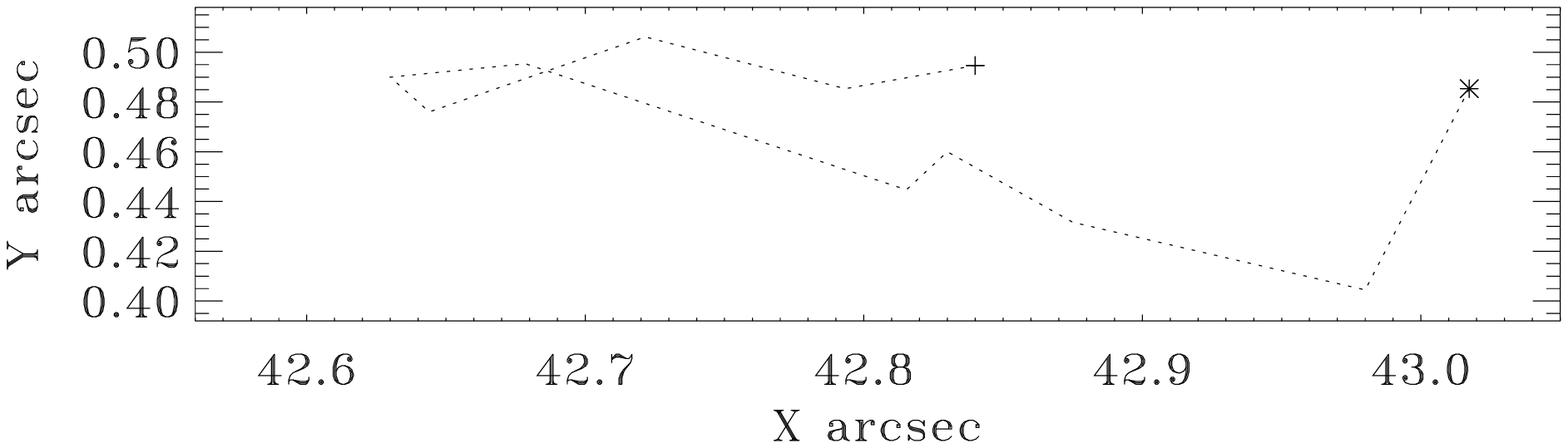}}
\caption[]{Trajectories of granules in a five-minute time sequence. Top:
Granule displacements in a small subfield. Bottom:
enlarged view of a granule trajectory. Data are from the 1988 Pic-du-Midi
series \cite[e.g.][]{MARVSFST92}.}
\label{traj}
\end{figure}

\subsection{Measuring the velocities}\label{vitesse}

Once the granules have been identified, the $(x,y)$ coordinates of their
barycentre are computed. Hence, each image is converted into a set of
points $\vX_{i,n}$ describing the position of the granules at time
$t_n$. These data may be completed by the set of granules surfaces,
shapes, etc.

The set of points $\{\vX_{i,n}\}$ is then divided into trajectories

\[ \left\{ \vX_{i(k),n}\right\}_{n_i \leq n\leq n_\ell}.\]
This notation means that granules of index $i(k)$ are in fact the same
granule as the one that follows the $k^{\rm th}$ trajectory; it appears at time
$t(n_i)$ and disappears at time $t(n_\ell)$.

Trajectories are identified by comparing each position $\vX_i$ on two
consecutive images and putting nearest neighbours together provided
their position does not differ more than a given threshold that is
determined by an upper bound on the velocity. Typically, we reject
velocities higher than 5~km/s.

If one disposes of a long time series of images, it may be useful to
determine the time evolution of the velocity field. For this purpose a
time window of width $\Delta t$ needs to be used and trajectories are
restricted to  time windows. Hence, for a given time window, one derives
a set of N trajectories:

\[ \left\{ \left\{ \vX_{i(k),n}\right\}_{n_i \leq n\leq n_\ell}^{k=1,N},
\, t\leq t(n_i),\,  t(n_\ell)\leq t+\Delta t\right\} \]

\noi existing in the field during the time $[t,t+\Delta t]$. We show
an example of such a set of trajectories in Fig.~\ref{traj}. We clearly see
from the enlarged view that granule motion is dominated by an erratic
motion that mixes (Earth) atmospheric noise with the turbulent random flow
of solar convection.

From this set we derive a mean velocity associated with each trajectory;
the $k^{\rm th}$ trajectory gives the velocity

\[ \moy{\vV_k} = \frac{\vX_{i(k),n_2} - \vX_{i(k),n_1}}{t(n_2)-t(n_1)},
\]
which we associate with the mean position of the trajectory

\[ \vX_k = \frac{1}{n_2-n_1+1}\sum_{n=n_1}^{n_2}\vX_{i(k),n}.  \]
Hence we end up with the set

\[ \left\{\vV_k, \vX_k\right\}_{k=1,N} \]
which describes the velocity field during the time interval $[t,t+\Delta
t]$.

\begin{figure}[th]
\centerline{\includegraphics[width=8cm]{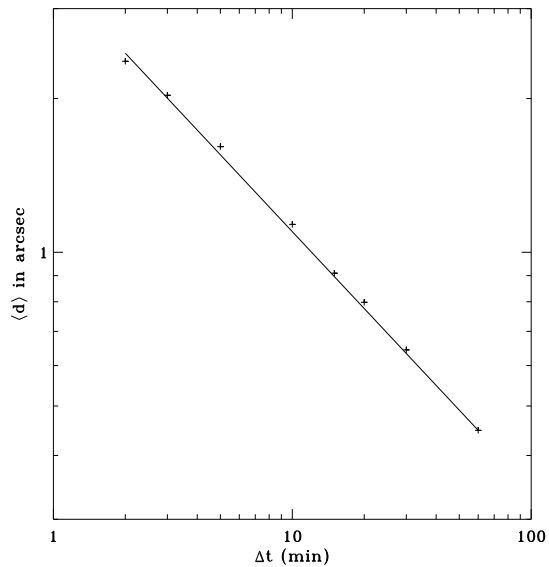}}
\caption[]{Mean distance between velocity vectors as a function of the
time window used for the measurement. The line shows the law $<d>\propto
1/\sqrt{\Delta t}$.}
\label{dens_pts}
\end{figure}

The values of the velocities are of course not uniformly distributed
in the field of view and we need to know how they constrain the velocity
field at a given resolution: small-scale components are weakly constrained,
while large-scale ones are highly constrained. The maximum resolution
for the velocity field is given by the density of trajectories and can
be estimated by the mean distance $<d>$ between the $\vX_k$. As the
time window $\Delta t$ is increased, the maximum resolution increases
according to the law $<d>\propto 1/\sqrt{\Delta t}$ as clearly shown
by Fig.~\ref{dens_pts}. This law arises because granules cover the Sun's
surface permanently. From it, we can derive the maximal spatial
resolution for a given time resolution. Indeed, from
Fig.~\ref{dens_pts}, it turns out that the mean volume of space-time
occupied by one granule is $\sim1200$~Mm$^2$s.

This value is useful for determining the highest time resolution that can
be allowed for the large-scale velocity fields. Indeed, as pointed out in
\cite{RRLNS01}, granules cannot be used to trace plasma flow under a
scale of 2.5~Mm (except in the case of very rapid flows like ``explosion'' of
granules); thus the determination of a large-scale flow needs a mesh
size not smaller than 1.25~Mm. This means that each grid point will have
a have a trajectory only if $\Delta t \geq 768s$. Conservatively, we
estimate that the time resolution cannot be higher than 15~min.

Often time resolution is not needed and therefore $\Delta t$ is much
larger than 15~min. In such a case,
several velocities may be given by granules appearing at a
given place on the grid. We then use the average velocity
of the granules as the measure of the local velocity. This introduces
another quantity, namely the rms fluctuation around this mean. This rms
velocity is a measure of the proper velocity of granules and therefore
a proxy of the local strength of turbulent convection.

\subsection{Derivation of the velocity field by MRA}

\begin{figure*}[htp]
\centerline{\includegraphics[width=18cm,angle=0]{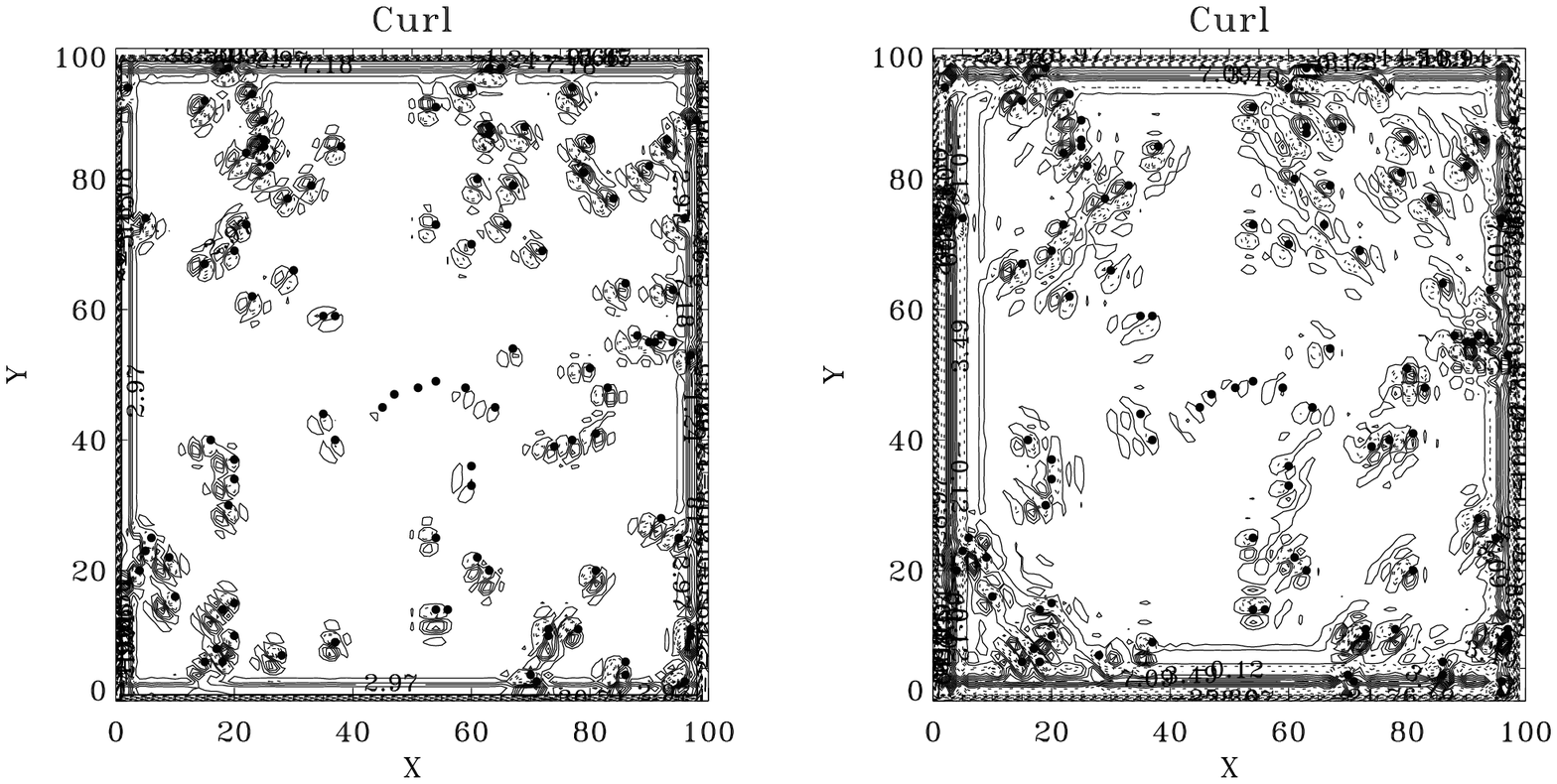}}
\caption[]{Isoline plot for the z-component of the curl of the velocity
field $v_x=-y, v_y=x$ with some values of $\vv$ randomly set to zero 
(black dots in the field). To the left, the scaling function $_4\phi$ of
Daubechies wavelet is used; to the right we use $_8\phi$. The main difference
between these two wavelets is the width of their support, twice larger
on the right. Note the border effects in both figures as well as the patterns
introduced by the absence of data and their dependence on the support of
the wavelet.  The large blank areas are at the constant value of 2, as
expected; solid lines represent isolines of a value different from 2 and
dotted lines show negative value isolines. X and Y units are grid points.}

\label{divtests}
\end{figure*}

\begin{figure*}[htp]
\centerline{\includegraphics[width=9cm,angle=0]{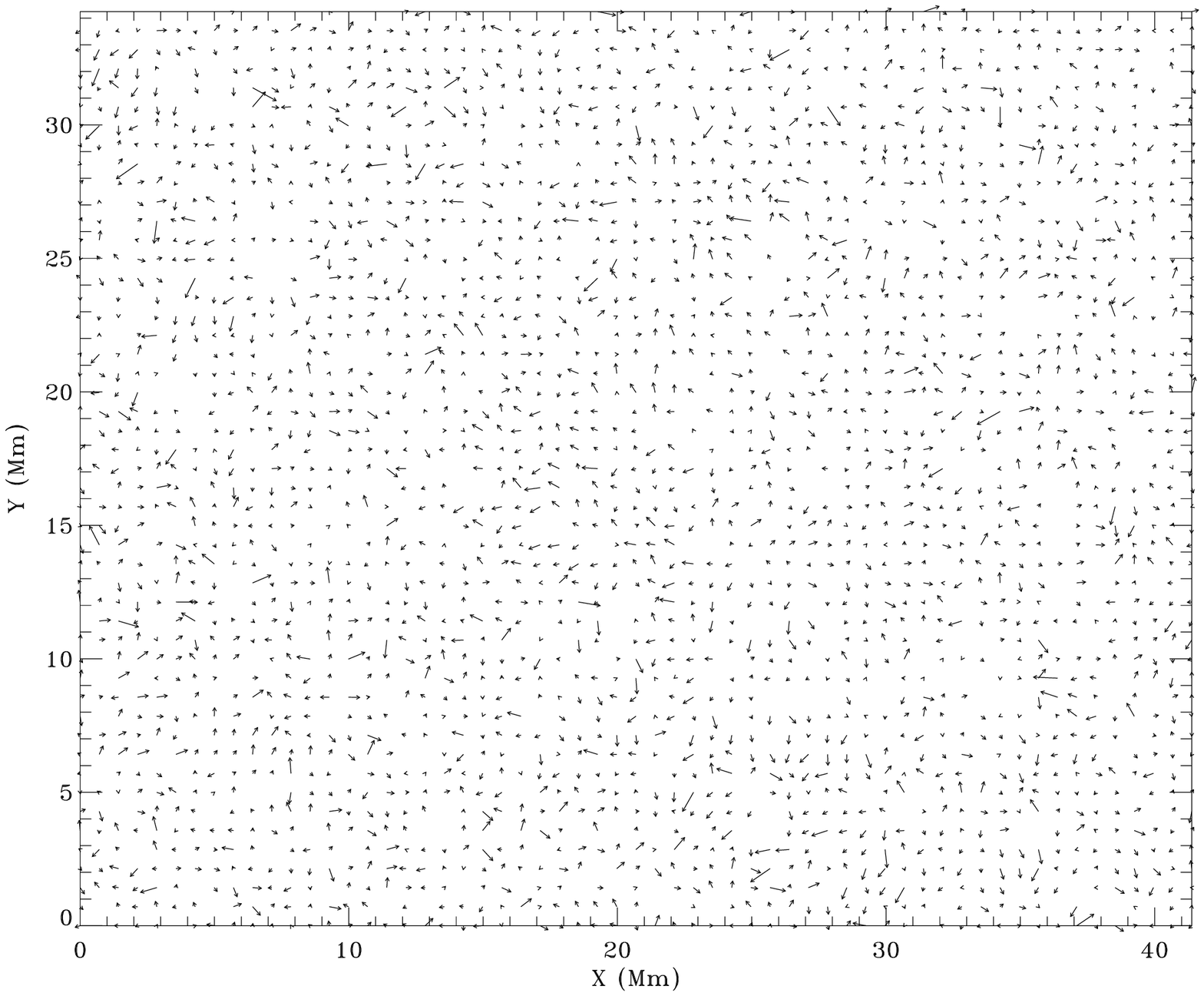}
\includegraphics[width=9cm,angle=0]{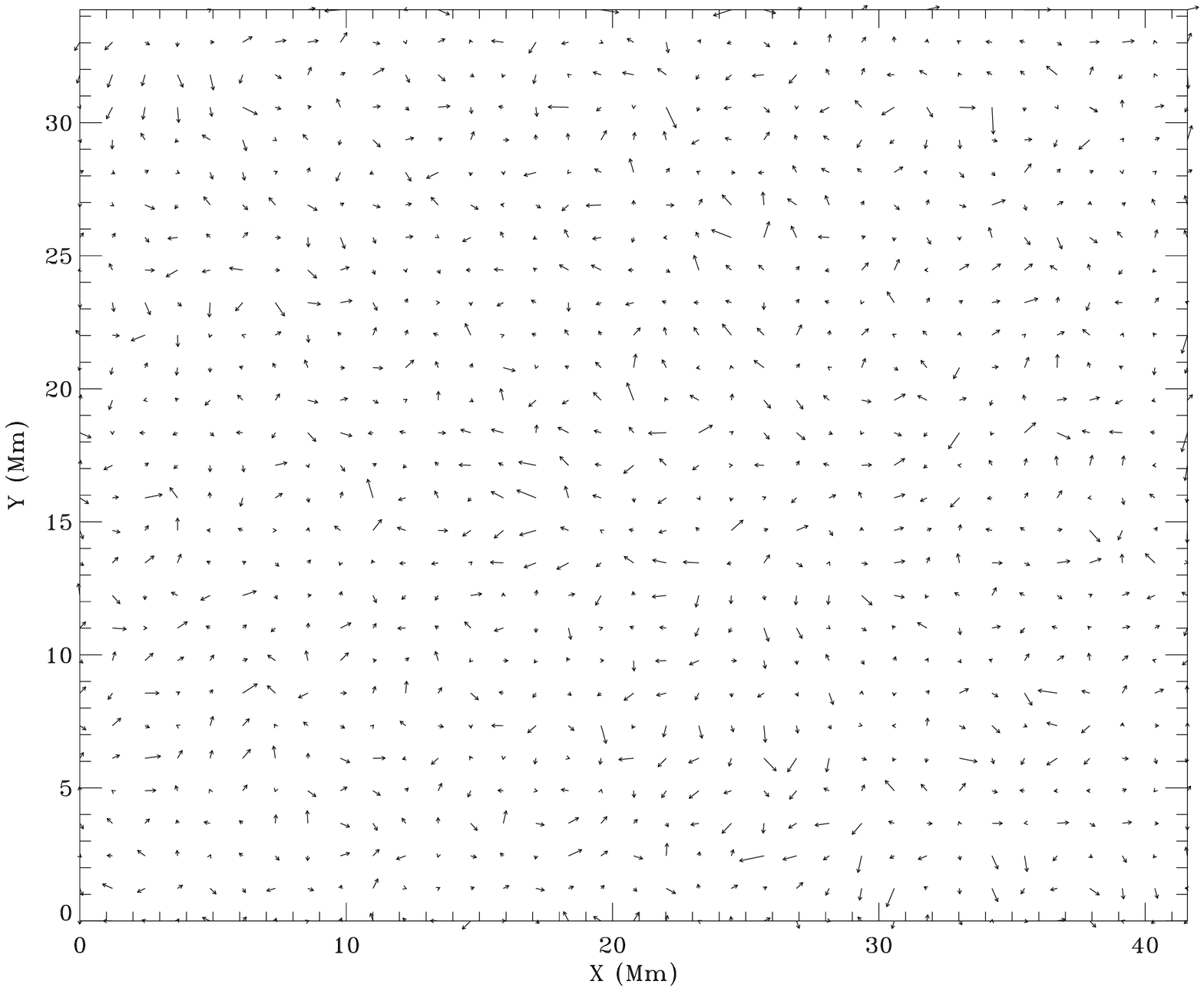}}
\caption[]{To the left, the velocity field with a small mesh size (713~km) is
complete with 83\% of data, while on the right, using a larger mesh size of
1223~km, the velocity field is complete to 99\%. Patterns of
velocities are easily identified between the two. We used a time
interval of 15~min.
}
\label{completeness}
\end{figure*}

Once the velocities in the field of view are determined, we need to know
what kind of flow they represent: for instance, vortices, shear layers,
diverging sources, etc. For this purpose we need to determine the best
continuous differentiable field that approximates the data. The
determination of such a field can be done in various ways, but we wish
to introduce no information  or a minimum of into this process. We also
wish to avoid the propagation of errors and noise in the field of view.

We have found that wavelet multi-resolution analysis (MRA) is an
interesting tool for this purpose, because it gives a decomposition of the
signal at all the scales allowed by the size of the box and therefore
gives a good view of all the components of the signal
\cite[]{MS89,MY87}.

The basic idea of a ``multi-resolution representation'' of $L^2(\bbbr)$
is to project a signal $ f $ on a ``wavelet orthonormal basis'' of
$ L^2({\bbbr}) $, at which point it is possible to extract the ``difference
in information'' between two successive approximations of the signal
(approximations at the resolution $ 2^j $ and $ 2^{j+1} $). The wavelet
orthonormal basis is a family of functions $\sqrt{2^j}\psi(2^jx-k)_{j,k\in
{\bbbz}} $ built by dilations and translations of a unique function
$\psi(x)$: the analysing wavelet.  The decomposition thus obtained is
this MRA. The signal can be
reconstructed from this representation without any difficulty. We give
a minimal background to this technique in appendix~\ref{appA}.

\begin{figure*}[htp]
\centerline{\includegraphics[width=9cm,angle=0]{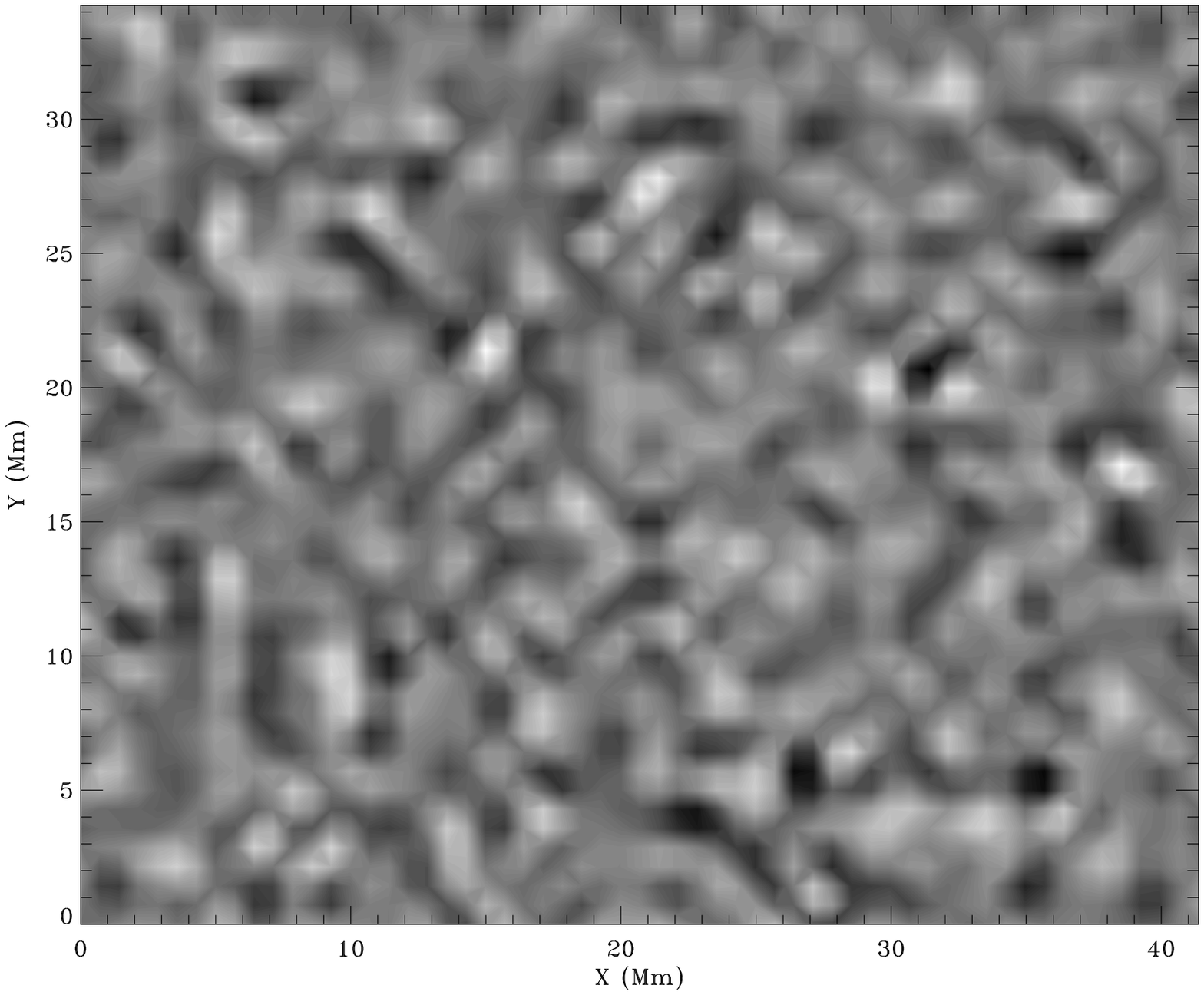}
\includegraphics[width=9cm,angle=0]{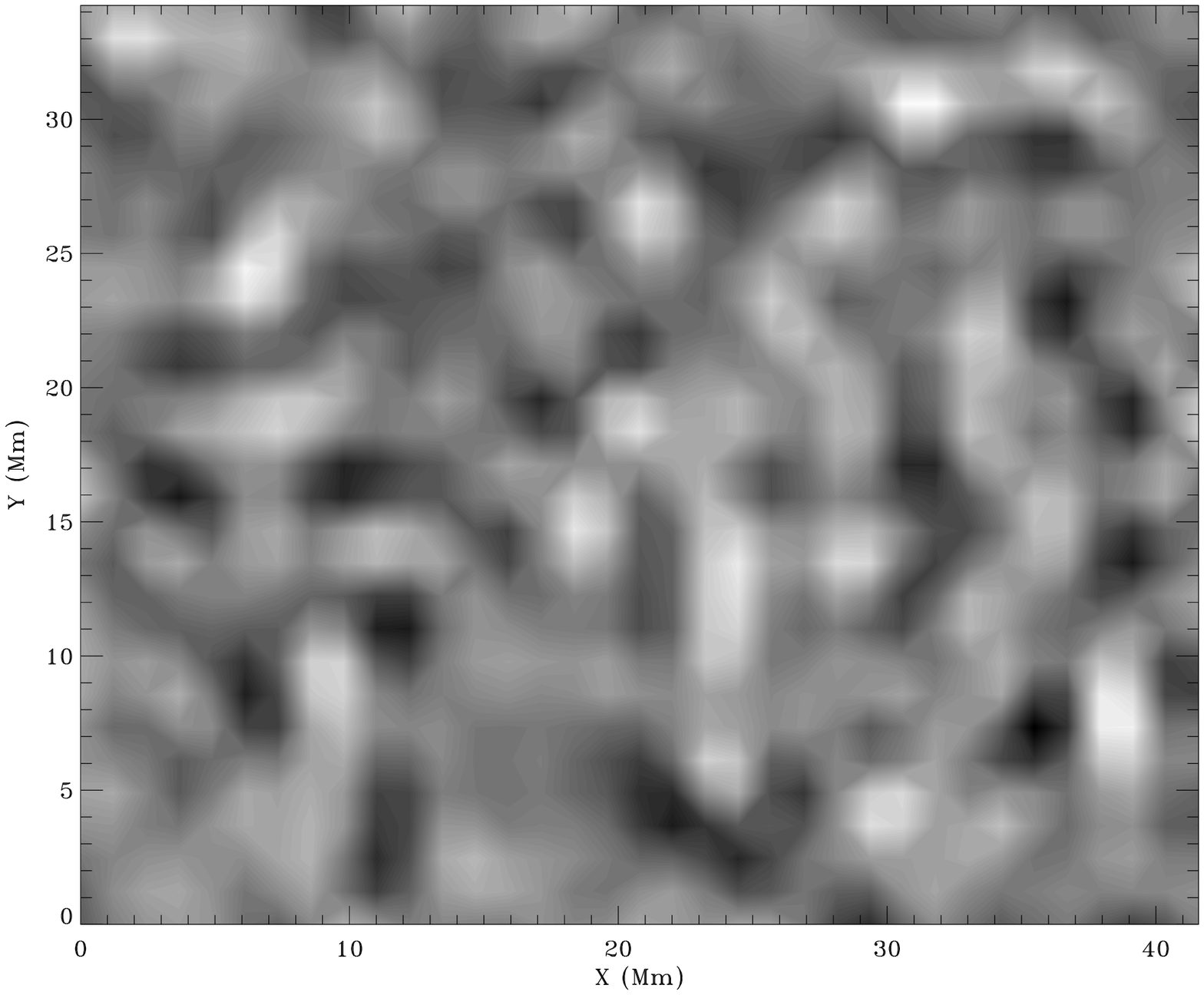}}
\centerline{\includegraphics[width=9cm,angle=0]{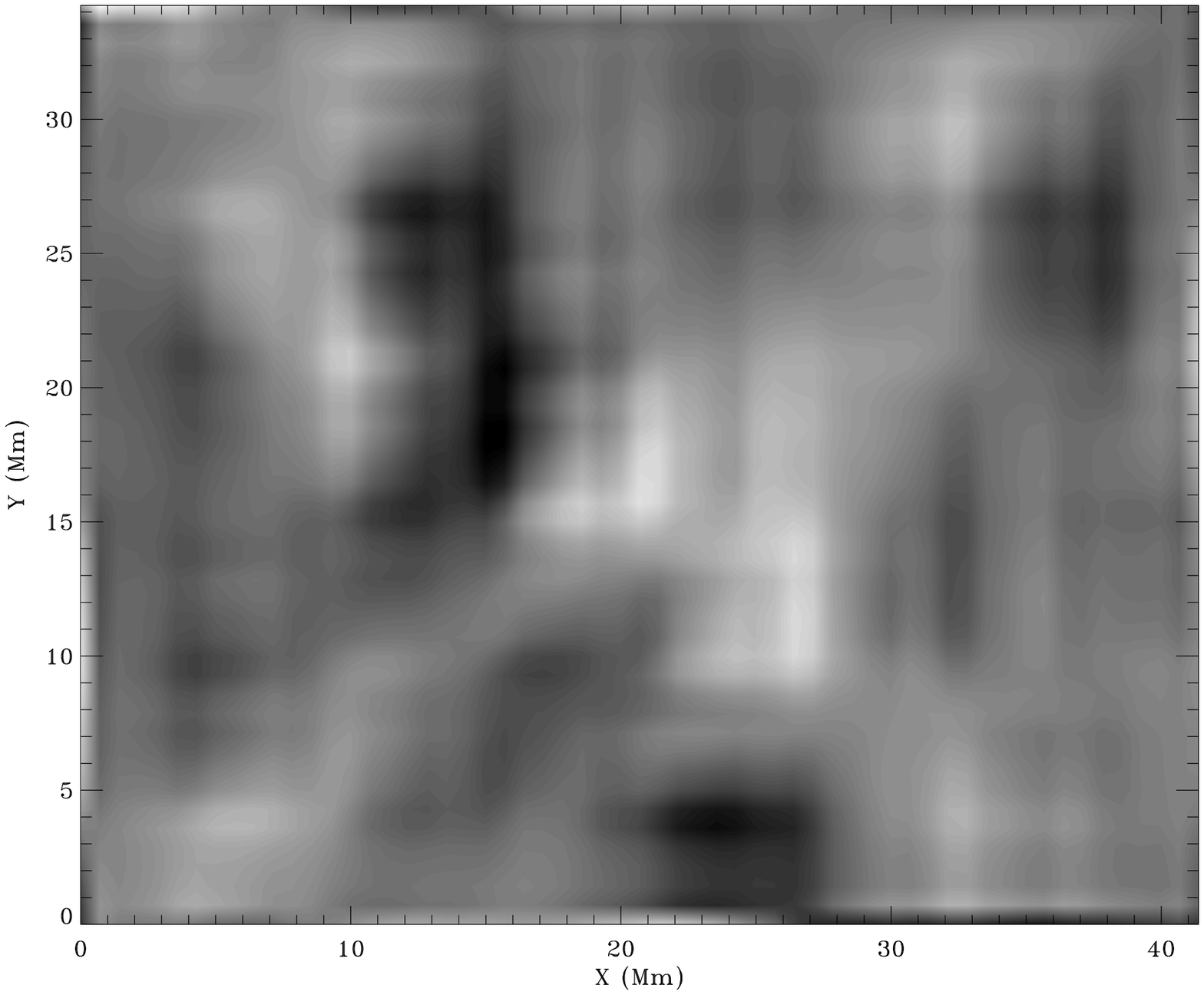}
\includegraphics[width=9cm,angle=0]{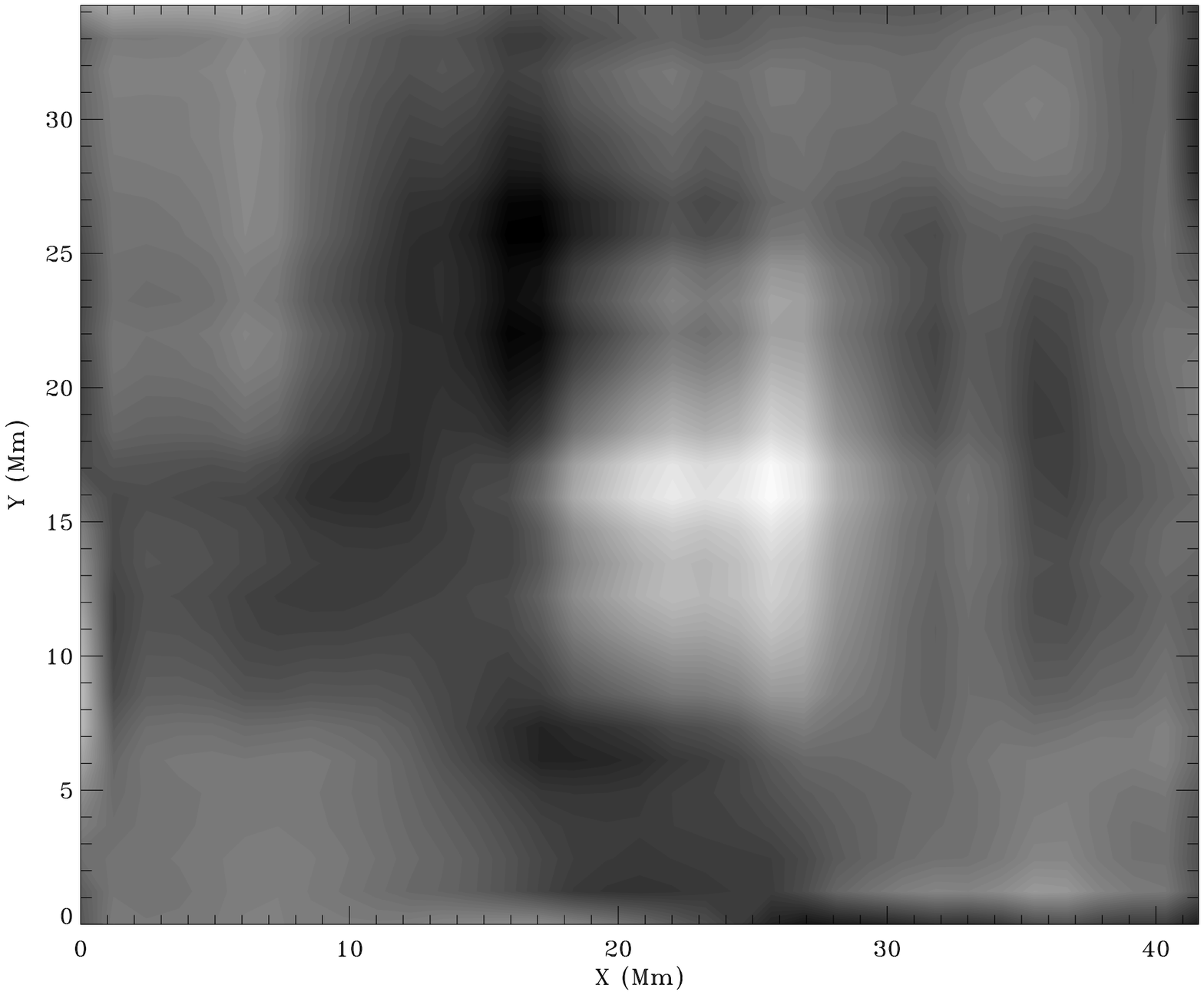}}
\caption[]{Divergence of the velocity field shown in
Fig.~\ref{completeness}. As in Fig.~\ref{completeness}, on the left we have
the more resolved, but less complete, sample and on the right the less-resolved,
but more complete, sample. In the first row the data have been
only slightly filtered (they are projected on the space j=1), and one
can hardly identify any common structure in the two plots. However, we
clearly see some common features between the two plots when considering
the low frequency j=3-component.}
\label{div_field}
\end{figure*}

Now we need to specify the choice of the analysing wavelet. As in many
problems of image processing, we choose the Daubechies wavelet
because of its compact support. This property is important since it
minimizes border effects and interactions between patterns of the signal
during the filtering process. Moreover, using these wavelets also
preserves the location of zero-crossing and maxima of the signal during
the analysis, a property that results in mutual suppressive
interactions across its different scale representations and superior
robustness in noisy environments \cite[]{SY93}. Thus, the contours of the
image can be determined efficiently.
We understand that this property  is important in image
processing since the features of the image are preserved after filtering.
For the velocity fields we are dealing with, this is also an interesting
point, because we wish to identify flow structures like divergences and
vortices.

Finally, wavelet analysis also allows us to determine the relevance (or
the reality) of flow structures on different scales. One may indeed
apply the MRA to the velocity field and the noise field. Then, for each
scale of the flow, we can compare the details and the amount of noise
to see whether the details are relevant or are simply noise structures.

\subsection{Curl and divergence fields}

Once an approximation of the velocity field is known, it is useful to
detect flow patterns that may be important for the dynamics of the
fluid. As the (measured) velocity field is purely two-dimensional, two
quantities are relevant for enhancing flow structures: the
divergence \mbox{$D=\partial_xv_x+\partial_yv_y$} and the $z$-component of the
vorticity \mbox{$\zeta=\partial_xv_y-\partial_yv_x$}.

The way derivatives can be computed can be explained with a
one-dimensional example. Let us consider the approximation on scale $j$
of the signal $f$

\beq f_j(x) = \sum_k \scal{f}{\phi_k^j} \phi_k^j(x), \eeqn{decompos}
where $k$ represents the position of the wavelet. Differentiating this
expression yields

\[ \dnx{f_j} = \sum_k \scal{f}{\phi_k^j} \dnx{\phi_k^j}. \]
In the Galerkin method, this formula would be sufficient; however, in MRA
the derivative of a function approximated with some resolution has
meaning only in the same functional space, that is, with the same
resolution. Therefore, $df_j/dx$ also needs to be projected onto that
space. Thus we are interested in

\[ \scalbig{\phi_n^j}{\dnx{\phi_k^j}}. \]
It therefore turns out that the discrete approximation of the
derivatives can be easily derived from that of the original functions by
a simple matrix multiplication. If $S_k^j=\scal{f}{\phi_k^j}$ is the
discrete approximation of $f$ on scale $j$ and $S'^j_k$ the discrete
approximation of $df/dx$ on the same scale, then it follows that

\beq S'^j_k = 2^{-j}\sum_l r_{k-l} S^j_l, \eeq
where

\[ r_l = \int_{-\infty}^{+\infty} \phi(x-l)\dnx{\phi}dx . \]
These numbers can be computed through an algorithm described in Beylkin
(1992).\nocite{Beylk92}

We show the computations of the curl field on a simple given velocity
field in Fig.~\ref{divtests}, namely $v_x=-y$, $v_y=x$. We see that,
except for the border effects due to the finite size of the filter, value
2 is correctly restored.

\subsection{The role of `holes'}

One of the problems arising when reconstructing the velocity field comes
from the presence of bins without data. Such bins  produce structures in
the divergence and curl fields. A simple illustration of the effect is
given in Fig.~\ref{divtests} where the curl of a solid rotation velocity
field is plotted. This figure shows the importance of the compact
support of the wavelet in limiting the propagation of errors.

Let us now consider some real data taken from the Pic du Midi
data set \cite[see][for details]{RRMV99}. Considering the velocity field
first, the effects of empty bins is not dramatic as may be seen in
Fig.~\ref{completeness}; velocity patterns are indeed not affected
much by holes. This is not the case for the divergence where we see, in
Fig.~\ref{div_field}, that when the number of empty bins is increased, the
patterns hardly remain identifiable. Only after a strong filtering can one
recover similar patterns.

The foregoing example illustrates what may be at the origin of the
worrying differences found by \cite{SBNSS95} between feature tracking and LCT,
namely an interpolation problem emphasised by considering
differentiated fields. The two methods will converge to similar results
only when these effects are overcome, which means that time or
space-averaging is large enough. Indeed, empty bins disappear either
when the sampling grid is coarse enough or when the time resolution is
low enough (see end of Sect.~\ref{vitesse}).

\section{Discussion and conclusions}

In this paper we have presented the coherent structure tracking algorithm
aimed at reconstructing the horizontal velocity field on the solar
surface from the granule motions.

We first discussed the role of segmentation and described a way
to generalise Strous algorithm. We have shown that a segmentation that
follows the shape of granules more closely is more sensitive to the
splitting of large granules and that, as far as velocity measurements
are concerned, the Strous criterion remains the most efficient.
We then showed that the reconstruction of the velocity field
is a delicate process because it is subject to many constraints. Indeed,
granules do not sample the field of view uniformly, and reconstruction of
the velocity field, along with its derivatives like divergence or curl,
requires some interpolation. There are many ways of performing such an
operation; however, classical methods, like polynomial interpolation,
would propagate errors and noise everywhere. We thus selected a method
based on MRA which projects data onto Daubechies
wavelets. The finite support of these functions limits the effects of
noise and error propagation: sides and regions lacking in data have a
limited influence. Moreover, the signal is decomposed at different
scales through a filtering process. At each step the filtered field and
the remaining details can be viewed and compared.This representation is
particularly  relevant for turbulent flows, since the relation between
amplitude  and scale is of crucial importance for constraining models of
these flows.

We also related the minimum size of the velocity mesh grid to the
time resolution. We thus found that, typically, one granule trajectory
occupies a ``volume" of 1200Mm$^2$s. When the ``space-time" resolution
does not reach this limit, many granules contribute to the velocity in
one mesh point. Their mean velocity is considered as the true local
velocity, but local fluctuations around this mean gives some information
on the local strength of convection.

We did not discuss here the influence of the Earth's atmospheric
turbulence on the determination of the velocity fields and derivatives.
Surely this is an important point: typical atmospheric distortion of
images induces, in good seeing conditions, feature motions of 0\farcs13.
For a long-lived structure, say 10~min, this means an uncertainty of
the velocity as high as 220~m/s. This is quite large compared to a
typical velocity of 600~m/s. Hence, atmospheric noise is a non negligible
part of the data and a careful determination of its influence is needed.
This is the subject of the companion paper to which we refer the reader 
\cite[see][]{TRMR06}.

Finally, although this has not been tested yet, we think that the CST
algorithm can be fruitfully used to track the magnetic features of the
photosphere, like network bright points or even determine velocities
of features in the solar atmosphere.

\appendix
\section{Fundamentals of multi-resolution analysis}\label{appA}

We give in this appendix the basic background of MRA and
refer the reader to textbooks for a more complete presentation
\cite[e.g.][]{Daub92,Mallat99}.  An MRA is a sequence $ {\lbrace V_j
\rbrace}_{j \in {\bf \bbbz}}  $ of closed subspaces of square-integrable
functions, $L^2(\bbbr)$, such that the five following properties are
satisfied, $ \forall j \in {\bbbz} $:

\begin{enumerate}
\item $ V_j \subset V_{j+1} $.

$ V_j $ can be interpreted as the set of all possible
signal approximations at the resolution $ 2^j $ (a resolution
$ r $ is defined by the size $ 1 / r $ of the smallest detail).
Thus, if the smallest detail in $ V_0 $ has size 1, it is
possible to read on $ V_j $ details of size $ 2^{-j} $.

It follows from this property that the approximation of
the signal at resolution $ 2^{j+1} $ contains all the
necessary information for determining the same signal at a lower
resolution $ 2^j $ (plus additional details).

\item $ \displaystyle{\bigcup _{j \in {\bbbz} }}V_j $
is dense in $ L^2(\bbbr) $ \quad and \quad $ \displaystyle
  {\bigcap _{j \in {\bf \bbbz} } }V_j = \lbrace 0 \rbrace $.

In other words, when $ j $ increases, the approximated
signal converges to the original signal. Conversely,
if $ j $ (the resolution) decreases, the approximated
signal converges to zero (it contains less and less
information).

\item $ f(x) \in V_j \Rightarrow f(2x) \in V_{j+1}$.

This property defines 2 as the rate of scaling change (the ratio
of two successive resolution values).

\item $ f(x) \in V_j \Rightarrow
    f(x-2^{-j}k) \in V_j \quad ,\quad \forall k \in {\bbbz} $.

This property characterizes the invariance under discrete
translations: when the signal is translated by a length
proportional to $ 2^{-j} $, the approximations are translated
by the same amount and no information is lost in the
translation.

\item A function $ \phi $ exists in
$ V_0 $ such that $ {\lbrace \phi(x-k) \rbrace} _{k \in {\bbbz}} $
is an orthonormal basis of $ V_0 $.
Hence, the family ${ \lbrace 2^{j /2} \phi (2^j x-k)
\rbrace}_{k \in {\bbbz}} $ is an orthonormal
basis of $ V_j $.
Function $ \phi $ is called the scaling function of
the multi-resolution representation.

\end{enumerate}
Now, let us define $ W_j $ as the orthogonal complement of $ V_j $
in $ V_{j+1} $ (it contains the additional details that
are in $ V_{j+1} $ and not in $ V_j $). There exists a function $ \psi $
(the wavelet)
such that $ \lbrace \psi(x-k) \rbrace _{k \in {\bbbz}} $
is an orthonormal basis of $ W_0 $ and
$ \lbrace 2^{j/2}\psi(2^j x -k) \rbrace _{j,k \in {\bbbz}} $
is an orthonormal basis of $ L^2(\bbbr) $.

One studies a signal $ f $ of $ L^2(\bbbr) $ by projecting it
orthogonally on the collection of $ V_j $ and $ W_j $.
This procedure can be carried out according to the pyramidal
algorithm presented below.

First, let us define the two filters $ h $
and $ g $ that can be deduced from the MRA.
This analysis allows us to determine a function $ h $,
which is the impulse response of some $ 2\pi $-periodic low-pass
filter $ H $ defined with the scaling function:
$ H(\omega)=\widehat{\phi}(2\omega)/\widehat{\phi}(\omega) $.
On the other hand, one defines function $ g $
by $ G(\omega)=\widehat{\psi}(2\omega)/\widehat{\phi}(\omega)  $,
$ g $ being the impulse response of the $ 2\pi $-periodic
high-pass filter $ G $. The filters $ H $ and $ G $ are
quadratic mirror filters and are linked by the relation
$ G(\omega)=e^{-i\omega}\overline{H(\omega+\pi)} $,
giving
$ g(n) = (-1)^{1-n}h(1-n) $ for the impulse responses. Then, the pyramidal architecture for
computing the wavelet representation can be easily written as:

\begin{itemize}

\item suppose that $ f(x_i) $ belongs to the $ V_0 $ ($ f(x_i)=f_0(x_i) $
approximation of the signal at resolution 1)
and decompose $ f(x_i) $ onto $ V_{-1}$
and $ W_{-1} $;

\item the decomposition onto $ V_{-1} $ consists in a convolution
by the filter $ \widetilde h, $ such that $ \widetilde h(n) = h(-n)$,
and a decimation (i.e.
only one out of every two sample is retained); we obtain the
$ N/2- $sampled so-called approximation
at resolution $ 1/2 $
equal to
  \[
f_{-1}(n)=\sum_{n=1}^N \widetilde h(2k-n)f_0(n) \in V_{-1}.
  \]

\item the decomposition onto $ W_{-1} $ consists in a convolution by
filter $ \widetilde g $ (such that $ \widetilde g(n)=g(-n) $)
and a decimation; we obtain in the same way
  \[
d_{-1}(n)= \sum_{n=1}^N \widetilde g(2k-n) f_0(n) \in W_{-1};
  \]
this is the detail at resolution $ 1/2 $, that is to say the
``difference in information'' between $ f_0(n) $ and
$ f_{-1}(n) $; it also has $ N/2 $ samples.
\end{itemize}
By repeating the same sequence, we obtain the approximation
and the detail at resolution $ 1/2^2 $ :
  \[ f_{-2}(n)= \sum_{n=1}^{N/2} \widetilde h(2k-n) f_{-1}(n)  \in V_{-2}
  \]
and
\[ d_{-2}(n)= \sum_{n=1}^{N/2} \widetilde g(2k-n) f_{-1}(n) \in
W_{-2} ,
  \]
and so on.

After a number $ J $ of iterations to be defined
by the problem, we have decomposed $ f_0 $ into
$ d_{-1},\ d_{-2},\ldots,d_J$, and $ f_J $.

Let us finally mention that
reconstruction of $ f_0(n) $ from the details
and the last approximation is just as easy and has appreciable
quality. One has to iterate (starting from $ j=J $) :
  $$
f_j(n)=2\Bigl( \sum_{k=1}^M h(n-2k) f_{j-1}(k) \allowbreak +
\sum_{k=1}^M g(n-2k) d_{j-1}(k) \Bigr),
  $$
$ f_{j-1} $ and $ d_{j-1} $ being sampled in $ M $ points.

The MRA can be generalised to two dimensions for image-processing
applications.  We can define a sequence of multi-resolution vector
spaces and the approximations of a signal $f(x,y) \in L^2(\bbbr^2)$
\cite[][]{MS89}.  Since the image under study is bounded, we choose $\phi$
with a compact support (\ie vanishes outside a finite region).  In that
case, filters $h$ and $g$ have only finitely many coefficients that
satisfy the MRA conditions \cite[][]{Daub92}.  Functions $\phi$ and
$\psi$ become more regular as the number of coefficients $n$ increases
(the case of $n=1$ corresponds to the discontinuous Haar basis).  It has been
proved \cite[][]{Daub92} that the regularity of $\phi$ and $\psi$ increases
linearly with $n$. By choosing $n=8$, we obtain a good compromise between
regularity and support width.


\bibliographystyle{aa}

\end{document}

%% file: 6491com.tex
\def\bbbr{{\rm I\!R}} 
\def\bbbc{{\mathchoice {\setbox0=\hbox{$\displaystyle\rm C$}\hbox{\hbox
to0pt{\kern0.4\wd0\vrule height0.9\ht0\hss}\box0}}
{\setbox0=\hbox{$\textstyle\rm C$}\hbox{\hbox
to0pt{\kern0.4\wd0\vrule height0.9\ht0\hss}\box0}}
{\setbox0=\hbox{$\scriptstyle\rm C$}\hbox{\hbox
to0pt{\kern0.4\wd0\vrule height0.9\ht0\hss}\box0}}
{\setbox0=\hbox{$\scriptscriptstyle\rm C$}\hbox{\hbox
to0pt{\kern0.4\wd0\vrule height0.9\ht0\hss}\box0}}}}
\def\bbbq{{\mathchoice {\setbox0=\hbox{$\displaystyle\rm
Q$}\hbox{\raise
0.15\ht0\hbox to0pt{\kern0.4\wd0\vrule height0.8\ht0\hss}\box0}}
{\setbox0=\hbox{$\textstyle\rm Q$}\hbox{\raise
0.15\ht0\hbox to0pt{\kern0.4\wd0\vrule height0.8\ht0\hss}\box0}}
{\setbox0=\hbox{$\scriptstyle\rm Q$}\hbox{\raise
0.15\ht0\hbox to0pt{\kern0.4\wd0\vrule height0.7\ht0\hss}\box0}}
{\setbox0=\hbox{$\scriptscriptstyle\rm Q$}\hbox{\raise
0.15\ht0\hbox to0pt{\kern0.4\wd0\vrule height0.7\ht0\hss}\box0}}}}
\def\bbbt{{\mathchoice {\setbox0=\hbox{$\displaystyle\rm
T$}\hbox{\hbox to0pt{\kern0.3\wd0\vrule height0.9\ht0\hss}\box0}}
{\setbox0=\hbox{$\textstyle\rm T$}\hbox{\hbox
to0pt{\kern0.3\wd0\vrule height0.9\ht0\hss}\box0}}
{\setbox0=\hbox{$\scriptstyle\rm T$}\hbox{\hbox
to0pt{\kern0.3\wd0\vrule height0.9\ht0\hss}\box0}}
{\setbox0=\hbox{$\scriptscriptstyle\rm T$}\hbox{\hbox
to0pt{\kern0.3\wd0\vrule height0.9\ht0\hss}\box0}}}}
\def\bbbz{{\mathchoice {\hbox{$\sf\textstyle Z\kern-0.4em Z$}}
{\hbox{$\sf\textstyle Z\kern-0.4em Z$}}
{\hbox{$\sf\scriptstyle Z\kern-0.3em Z$}}
{\hbox{$\sf\scriptscriptstyle Z\kern-0.2em Z$}}}}

\newcommand{\ie}{{\it i.e.}\ }

\newcommand{\beq}{\begin{equation}}
\newcommand{\beqa}{\begin{eqnarray*}}
\newcommand{\beqan}{\begin{eqnarray}}
\newcommand{\greq}{\begin{equation}\left\{ \begin{array}{l}}
\newcommand{\egreq}{\end{array}\right. \end{equation}}
\newcommand{\nngreq}{\[\left\{ \begin{array}{l}}
\newcommand{\nnegreq}{\end{array}\right. \]}
\newcommand{\egreqn}[1]{\end{array}\right. \label{#1}\end{equation}}
\newcommand{\eeq}{\end{equation}} 
\newcommand{\eeqn}[1]{\label{#1}\end{equation}} 
\newcommand{\eeqa}{\end{eqnarray*}}
\newcommand{\eeqan}[1]{\label{#1}\end{eqnarray}}

\newcommand{\noi}{ \noindent }

\newcommand{\vv}{\vec{v}}
\newcommand{\vV}{\vec{V}}

\newcommand{\vX}{\vec{X}}

\newcommand{\dnx}[1]{\frac{d  #1}{dx}}

\newcommand{\moy}[1]{\left\langle #1\right\rangle}

\newcommand{\scal}[2]{\bigl\langle #1\bigr| #2 \bigr\rangle\,}
\newcommand{\scalbig}[2]{\Bigl\langle #1\Bigr| #2 \Bigr\rangle\,}

